\documentclass[titlepage,12pt]{article}
\usepackage{amssymb,psfig,epsfig,pslatex}
\usepackage{axodraw}

\usepackage{cite}
\def\@sect#1#2#3#4#5#6[#7]#8{\ifnum #2>\c@secnumdepth
  \def\@svsec{}\else
  \refstepcounter{#1}\edef\@svsec{\csname the#1\endcsname.\hskip0.5em}\fi
  \@tempskipa #5\relax
  \ifdim \@tempskipa>\z@
    \begingroup
      #6\relax
      \@hangfrom{\hskip #3\relax\@svsec}{\interlinepenalty \@M #8\par}%
    \endgroup
    \csname #1mark\endcsname{#7}\addcontentsline
      {toc}{#1}{\ifnum #2>\c@secnumdepth \else
        \protect\numberline{\csname the#1\endcsname}\fi #7}%
  \else
    \def\@svsechd{#6\hskip #3\@svsec #8\csname #1mark\endcsname
      {#7}\addcontentsline{toc}{#1}{\ifnum #2>\c@secnumdepth \else
        \protect\numberline{\csname the#1\endcsname}\fi #7}}%
  \fi \@xsect{#5}}

\usepackage{color}
\begin{document}

\begin{center}
{\LARGE {\bf Colour Connection   and  Diquark Fragmentation in 
$e^+e^- \to c \bar c q \bar q \to h's$ Process }}\\
\vspace{1.5cm}
{\large Wei Han\footnote{hanwei@mail.sdu.edu.cn}\,$^a$, 
Shi-Yuan Li\footnote{lishy@sdu.edu.cn}\,$^{a,b}$, 
Zong-Guo Si\footnote{zgsi@sdu.edu.cn}\,$^a$, Zhong-Juan
Yang\footnote{yangzhongjuan@mail.sdu.edu.cn}\,$^a$}\\
\par\vspace{1cm}
$^a$ Department of Physics, Shandong University, Jinan, 250100, 
P. R. China\\
$^b$ Institute of Particle Physics, Huazhong Normal University, Wuhan, 
430079, P. R. China


\par\vspace{1cm}
{\bf Abstract}
\parbox[t]{\textwidth}

\end{center}

\vspace{0.1cm}

 The hadronization effects 
induced by different colour connections of 
$c\bar{c}q\bar{q}$ system in $e^+e^-$ annihilation
 are investigated
by a toy model where diquark fragmentation is employed based on
Pythia. It is found that the  correlations  between the 
charm baryons and charm antibaryons produced via 
diquark pair fragmentation
 are much stronger, 
and their  momentum  spectra are  harder than those from  
the standard colour  connection in Pythia.

\vspace{0.3cm}
{\it keywords:} Colour Connection, Diquark, Hadronization

\vspace{1cm}

The hadronization  for partons fragmenting into hadrons is   
very important, and can only be treated by 
hadronization models up to now. 
The colour connection plays a crucial  r\^{o}le to set  the ``surface''  
between  the Perturbative Quantum Chromodynamics (PQCD) phase and  the
hadronization one, while 
the  topic related to the colour connection is beyond the approach of PQCD. 
It is argued  that some clews can be drawn 
from analyzing the decomposition of the colour  
space of the final partons produced 
in  PQCD phase \cite{Gustafson,xie1,xie2}.
For  the colour space of  the  parton system, 
there are many decomposition ways  corresponding to various 
colour connections, respectively (see, e.g., \cite{WQEFH,Wagus}). 
Hence, it is instructive to study the 
special properties of the final hadrons, which could 
give  information for the evidence of  
specific colour connections. Such kinds of investigation have been 
made for various parton systems \cite{Friberg:1996xc,xie3,LEP2}. 
One   interesting and  important  example 
is the    ($q_1$ $\bar q_2$ $q_3$ $\bar q_4$) system
produced via hard collisions. This is the simplest  system where many 
phenomena/mechanisms  related to QCD properties 
could be studied, e.g.,  (re)combination of quarks 
in producing special hadrons,  influence on  
reconstruction of intermediate particles (like
$W^\pm$) by soft interaction, etc.,
 most of which  more or less relate with the colour connections among the 
 four (anti)quarks. In this letter,  the
possible colour connections of this system are briefly reviewed.  
It is pointed out  that  the hadronization effects of 
one kind  of the colour connections  have never been modeled yet. Its 
special case (diquark pair fragmentation) in $e^+ e^-$ annihilation
is investigated in detail.

 For   $q_1 \bar q_2 q_3 \bar q_4$
system, two of the 
decomposition ways of its colour 
space are given as follows:
\begin{equation}
\label{1stdp}
(3_1 \otimes 3^*_2) \otimes (3_3 \otimes 3^*_4)= 
(1_{12} \oplus 8_{12}) \otimes (1_{34} \oplus
 8_{34}) = (1_{12} \otimes 1_{34}) \oplus 
 (8_{12} \otimes 8_{34}) \oplus \cdots,  
\end{equation}
\begin{equation}
\label{2nddp}
(3_1 \otimes 3^*_4) \otimes (3_3 \otimes 3^*_2)= ( 1_{14} \oplus 8_{14}) 
\otimes ( 1_{32} \oplus  8_{32})=(1_{14} \otimes 1_{32}) \oplus (8_{14}\otimes 
8_{32}) \oplus \cdots, 
\end{equation}
where  $3$, $3^*$, $1$, $8$  respectively
denote the representations triplet, anti-triplet,
singlet and  octet of the Group $SU_c(3)$, and  
the subscripts  correspond to the relevant (anti)quarks. 
In these two colour decompositions,
quark and antiquark form a cluster/string which fragments into 
hadrons independently. This picture is adopted in the 
popular hadronization models (for detail, see 
\cite{string,pythia,cluster,herwig}). The difference between these 
two decompositions leads to the colour reconnection phenomena as 
discussed in the processes 
 $e^+e^- \to W^+ W^-/Z^0 Z^0 \to q_1 \bar q_2 q_3 \bar q_4 \to 4~ jets $
 at LEPII\cite{LEP2, khojos, gustafsonw, Gustafson:1994cd,xie1,lsya}, etc.
One might also notice, in  the  $e^+ e^- \to 
q_1 \bar q_1+g^* \to
q_1 \bar q_1 q_2 \bar q_2$ process, 
($q_1 \bar q_1$) and ($q_2 \bar q_2$)
cannot be in colour-singlets, while
 ($q_1 \bar q_2$) and ($q_2 \bar q_1$)
are usually treated as colour-singlets and  hadronize independently. 
This corresponds to the term $1_{14}\otimes 1_{32}$ in eq.
(\ref{2nddp}) \cite{xie4}.

\begin{figure}
\begin{center}
\fcolorbox{white}{white}{
\begin{picture}(500,200) (34,-35)
\SetWidth{0.7}
\SetColor{Black}
\ArrowLine(133,42)(211,43)
\DashArrowLine(100,41)(142,67){2}
\ArrowLine(142,67)(187,100)
\ArrowLine(194,82)(142,67)
\ArrowLine(211,13)(40,12)
\GOval(210,27.5)(16,6)(0){0.99}
\GOval(192,92)(10,8)(0){0.99}
\Text(203,90)[lb]{\Large{\Black{$\pi^-$}}}
\Text(221,29)[lb]{\Large{\Black{$D^0$}}}
\Text(115,-40)[lb]{(a)}
\ArrowLine(40,42)(133,42)
\GOval(39,27)(14,8)(0){0.99}
\Text(13,26)[lb]{\Large{\Black{$B^-$}}}
\ArrowLine(300,49)(381,49)
\DashArrowLine(381,49)(431,26){2}
\ArrowLine(431,26)(463,79)
\ArrowLine(463,0)(431,26)
\ArrowLine(465,-15)(300,-15)
\ArrowArcn(516,37)(59,225,142)
\GOval(469,86)(20,11)(0){0.99}
\ArrowLine(381,49)(465,105)
\GOval(469,-3.6)(12,6)(0){0.99}
\GOval(300,16)(33,12.5)(0){0.99}
\Text(67,51)[lb]{\Large{\Black{$b$}}}
\Text(167,51)[lb]{\Large{\Black{$c$}}}
\Text(100,53)[lb]{\Large{\Black{$W^-$}}}
\Text(67,19)[lb]{\Large{\Black{$\bar{u}$}}}
\Text(271,18)[lb]{\Large{\Black{$B^-$}}}
\Text(330,59)[lb]{\Large{\Black{$b$}}}
\Text(410,83)[lb]{\Large{\Black{$c$}}}
\Text(411,38)[lb]{\Large{\Black{$W^-$}}}
\Text(330,-5)[lb]{\Large{\Black{$\bar{u}$}}}
\Text(486,86)[lb]{\Large{\Black{$\Sigma^0_c$}}}
\Text(480,-12)[lb]{\Large{\Black{$\bar p$}}}
\Text(380,-40)[lb]{(b)}
\end{picture}
}
\end{center}
\caption{ (a) $B^- \to \pi^-+ D^0$ and (b) 
$B^- \to \bar{p}+\Sigma_c^0$ processes.}
\label{bde}
\end{figure}
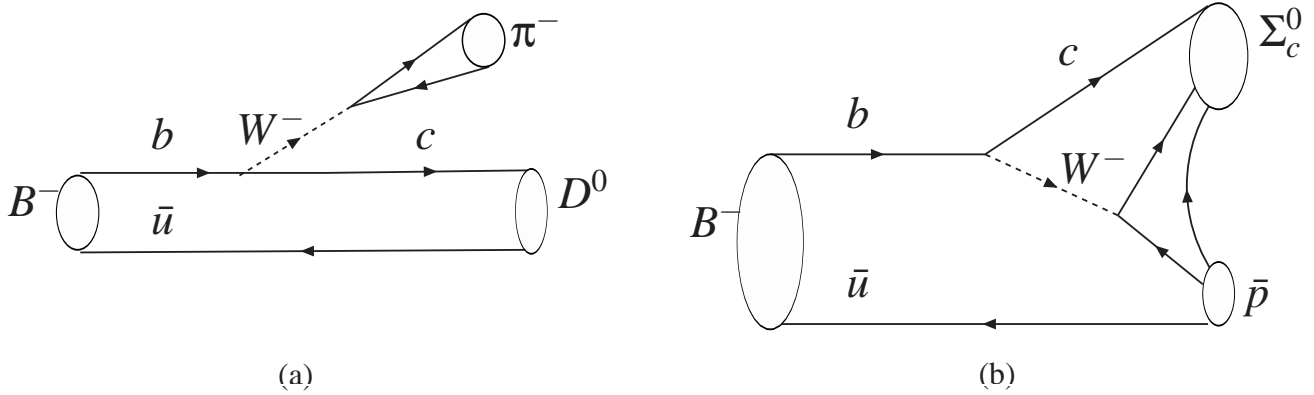

The colour space of $q_1 \bar q_2 q_3 \bar q_4$ system can also be
decomposed as follows:
 \begin{equation}
(3_1\otimes 3_3) \otimes (3^*_2 \otimes 3^*_4) = 
(3^*_{13} \oplus 6_{13}) \otimes 
(3_{24} \oplus 6^*_{24})=(3^*_{13}\otimes 3_{24})
\oplus (6_{13}\otimes 6^*_{24})
 \oplus \cdots, 
\label{diqcl}
\end{equation}
where $6_{13}$ ($6^*_{24}$) denotes the representation sextet (anti-sextet)
of the Group $SU_c(3)$.
A special case is that 
two quarks in colour state 3*  
attract each other and form a ``diquark''  when their
 invariant mass is small enough, and similarly
two antiquarks form an antidiquark. 
For this decomposition, only the whole four (anti)quarks system can form
a colour singlet,  hence they must 
hadronize altogether.
 This hadronization picture is 
quite different from that adopted in the popular hadronization models.
Unfortunately, this picture has not been taken into account
 though it 
is crucial and exists indeed in many processes.
One example is 
the  exclusive $B^-$ decay as shown in fig. \ref{bde}
where $b\to c + W^{-*} \to c+ d+\bar{u}_1$.
In this process,
the quark-antiquark pairs $d \bar{u}_1$ 
and $c \bar u$
are in colour-singlet,
and naturally form two mesons  $\pi^-$ and $D_0$ (fig.\ref{bde}(a)).
Here the combination for $d \bar{u}$ 
and $c \bar{u}_1$ is also possible. These two combinations 
correspond to the colour connections shown in eqs. 
(\ref{1stdp}) and (\ref{2nddp}), respectively.
The branching ratio of $B^- \to \pi^- D^0$ has been measured 
to be  $(4.98 \pm 0.29) \times 10^{-3}$ \cite{pdg}.
 In the meantime,
it is very interesting to notice that another kind of colour 
connection corresponding to eq. (\ref{diqcl}) exists, i.e., 
the $cd$ and $\bar{u}\bar{u}_1$ are 
in colour state $3^*$ and $3$, and combine with a 
$d$ or $\bar{d}$ from  vacuum to form two baryons $\Sigma^0_c$ and $\bar p$,
respectively. 
This is a  `signal channel' for diquark-antidiquark pair 
fragmentation, and its 
branching ratio is 
$(0.45^{+0.26}_{-0.19} \pm 0.007 \pm 0.12) \times 10^{-4}$ \cite{btsigma}. 
Additionally, the  three body baryonic decay processes  such as 
$B^- \to \Lambda_c \bar p \pi^-$ 
($Br=(2.1 \pm 0.7)\times 10^{-4} $ \cite{pdg}) can also be regarded to be 
induced by diquark fragmentation.
 Another example is the doubly-heavy baryon production
in high energy collisions\cite{ma2003,chang06}.
Recently, SELEX Collaboration observed the hadronic 
production of 
 $\Xi_{cc}$ at Fermilab \cite{mattson}.  This is  dominated by the 
production of cc pair in   PQCD process, 
and can  also  give an unambiguous confirmation of the existence of the 
diquark fragmentation. 

The aim of this letter is to investigate the
diquark fragmentation in $e^+e^-\to q_1\bar{q}_1 q_2\bar{q}_2\to h's $ process.
The accumulated/accumulating  data in various $e^+e^-$ colliders
can provide  sufficient statistics 
to study the hadronization of the four quark
states. 
In order to specify different colour connections 
of the four quark system via 
measuring  the corresponding hadron(s),
we focus on the  $e^+e^- \to c \bar c q \bar q \to h's$ process, 
where $q$ represents  $u$, $d$, or $s$ quark.
 The {\it ensemble} properties related to a class of hadrons (e.g.,
charm baryons) are investigated, so that  
we can   search for  the quantities 
which could be  measured in experiments.  
The statistics can also be enhanced by studying a group of hadrons.

As mentioned above,  the  hadronization
of the colour connection displayed in eq. (\ref{diqcl}) 
has not been considered in popular hadronization models.
However, 
one can employ the  
event generator Pythia\cite{pythia} based on Lund model\cite{string},
to describe the hadronization of the special case
that both $cq$ (in $3^*$) and  $\bar{c}\bar{q}$ (in 3) 
have small invariant masses so that they can be regarded
as diquark and antidiquark, respectively. 
For this specific case, there are 
strong constrains on both colour state and phase space.
Therefore it is important to investigate the 
fraction of this specific case, and 
relevant properties, at parton level.

\begin{figure}
\begin{center}
\psfig{file=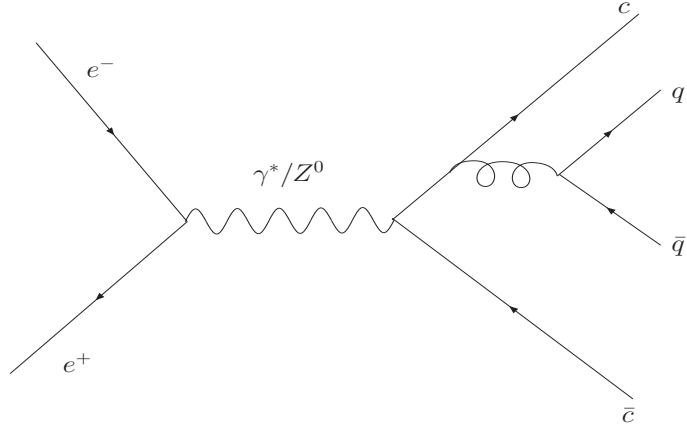, height=6cm, width=10cm}
\caption{One of the four Feynman diagrams for $e^+e^- \to c \bar c q \bar q$ 
process.}
\label{fey4q}
\end{center}
\end{figure}

\begin{figure}
\begin{center}
\psfig{file=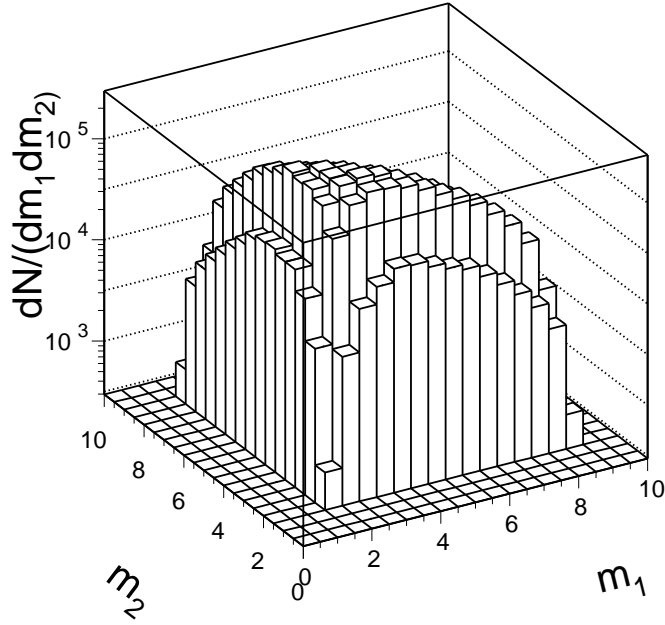, width=10cm}
\caption{The joint distribution $\frac{d^2 N}{d m_1 d m_2}$ 
for $e^+e^- \to c\bar{c} u\bar{u}$.}
\label{massd}
\end{center}
\end{figure}

{\large \it Parton level analysis}~~
At leading order QCD, the differential cross section for 
 $e^+e^- \to c \bar c q \bar q$  process(fig. \ref{fey4q})
can be written as 
\begin{equation}
\label{cr}
d \sigma= \frac{1}{2s} ~ d\, {\cal L}ips_4 ~ \overline{|M|^2},
\end{equation}
where  $s$ is the total energy square, ${\cal L}ips_4$ represents
 the 4-particle phase space
and $\overline{|M|^2}$ is the spin averaged-summed invariant 
amplitude square. 
At the energy of B factories (eg., $\sqrt{s}=10.52 GeV$),
we calculate the joint distribution 
 ${d^2 N}/(d m_1 d m_2)$ where $m_1$ and $m_2$ respectively 
represent the invariant
mass of $cq$ and $\bar{c}\bar{q}$.
In our calculation, we take $q=u$ as an example, and 
set the precise structure 
constant to be  ${1}/{137}$,
the strong coupling constant $\alpha_s=0.2$, the quark masses $m_c=1.5
GeV/c^2$ 
and $m_{u,d}=0.33 GeV/c^2$. 
The corresponding results are 
shown in  fig. \ref{massd}.
It is important to notice that the production probability is not
quite small when $m_1$ and $m_2$ are close to 
the diquark mass\footnote{The diquark mass  is around $2 GeV/c^2$
 set by  Pythia.}.
Especially, when the luminosity is large, the absolute event number
with diquark pair production could be large enough to be measured. 
Employing the KEK B factory integrated luminosity 
($525.828 fb^{-1}$ until Dec. 23, 2005) we obtain
the number of the events with $m_1, \, m_2 \leq 2.5$ $GeV/c^2$
to be around $3 \times 10^4$.
The number of events for $q=d$ or $q=s$
  is  at the same order.

\begin{figure}
\begin{center}
\psfig{figure=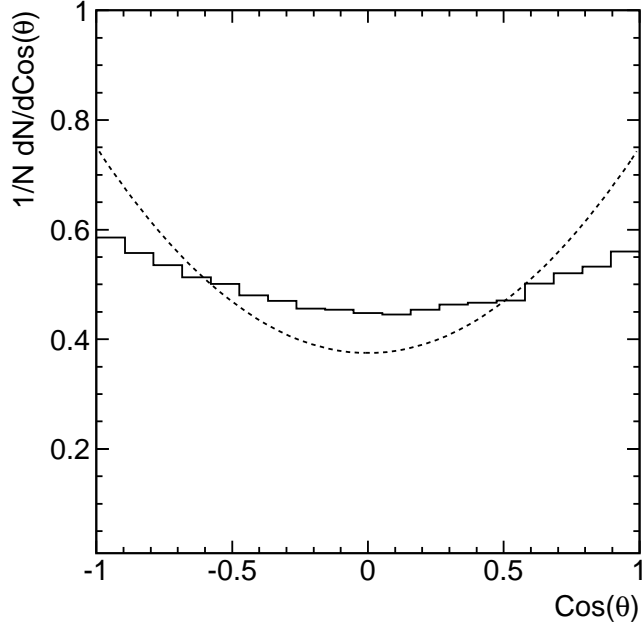, width=10cm}
\caption{Normalized polar angle distribution of diquark $cq$ 
(solid histogram). 
The dashed curve is for the normalized  $1+cos^2 \theta$ distribution.}
\label{thds}
\end{center}
\end{figure}

It is well known that in the  $e^+e^- \to \gamma^* \to f \bar f$ process,
the polar angle distribution of the massless spin $1/2$ fermion ($f$/$\bar f$) 
is predicted to be
 $1+cos^2 \theta$. For the  2-jet events initialized  by 
$e^+ e^- \to q \bar q$ at energies much smaller than 
$M_{Z^0}$, this will be approximately reproduced for the local parton hadron 
duality.
For the two clusters $cq$ and  $\overline{cq}$ fragmentation, the 
event shape is very close to 2-jet event.
If the polar angle distribution in this case is much 
different from the lowest order one, 
it can be expected as a possible signal to identify diquark pair fragmentation
at B factory energy.
The corresponding distributions with $m_1,\, m_2\leq 2.5GeV$  
are shown in fig. \ref{thds}. 
Obviously the distribution is quite different from $1+cos^2 \theta$, where
$\theta$  refers to the angle between 
the cluster $cq$  momentum and the $e^-$ momentum.

\begin{figure}
\begin{center}
\psfig{figure=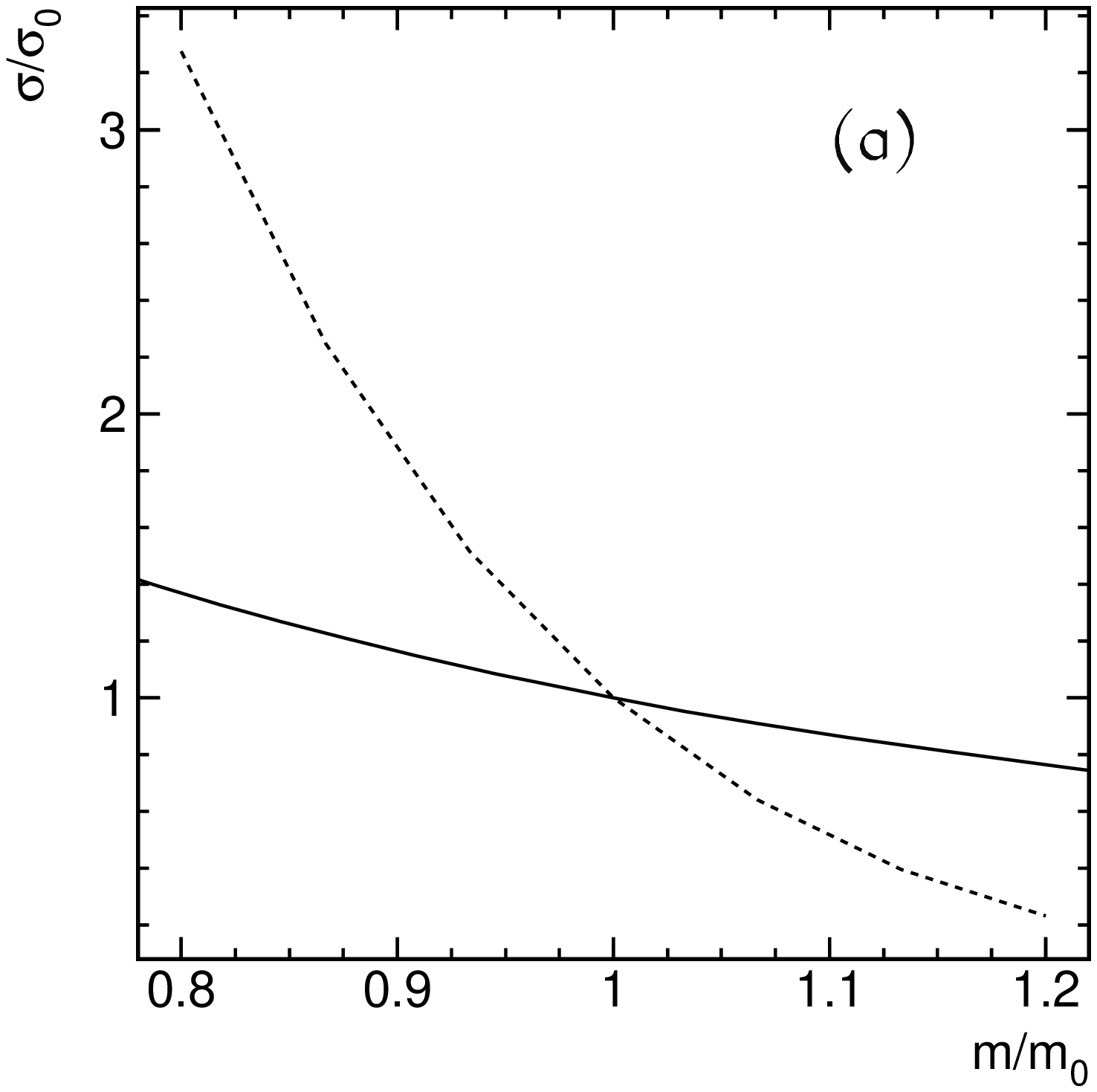, width=8cm} \psfig{figure=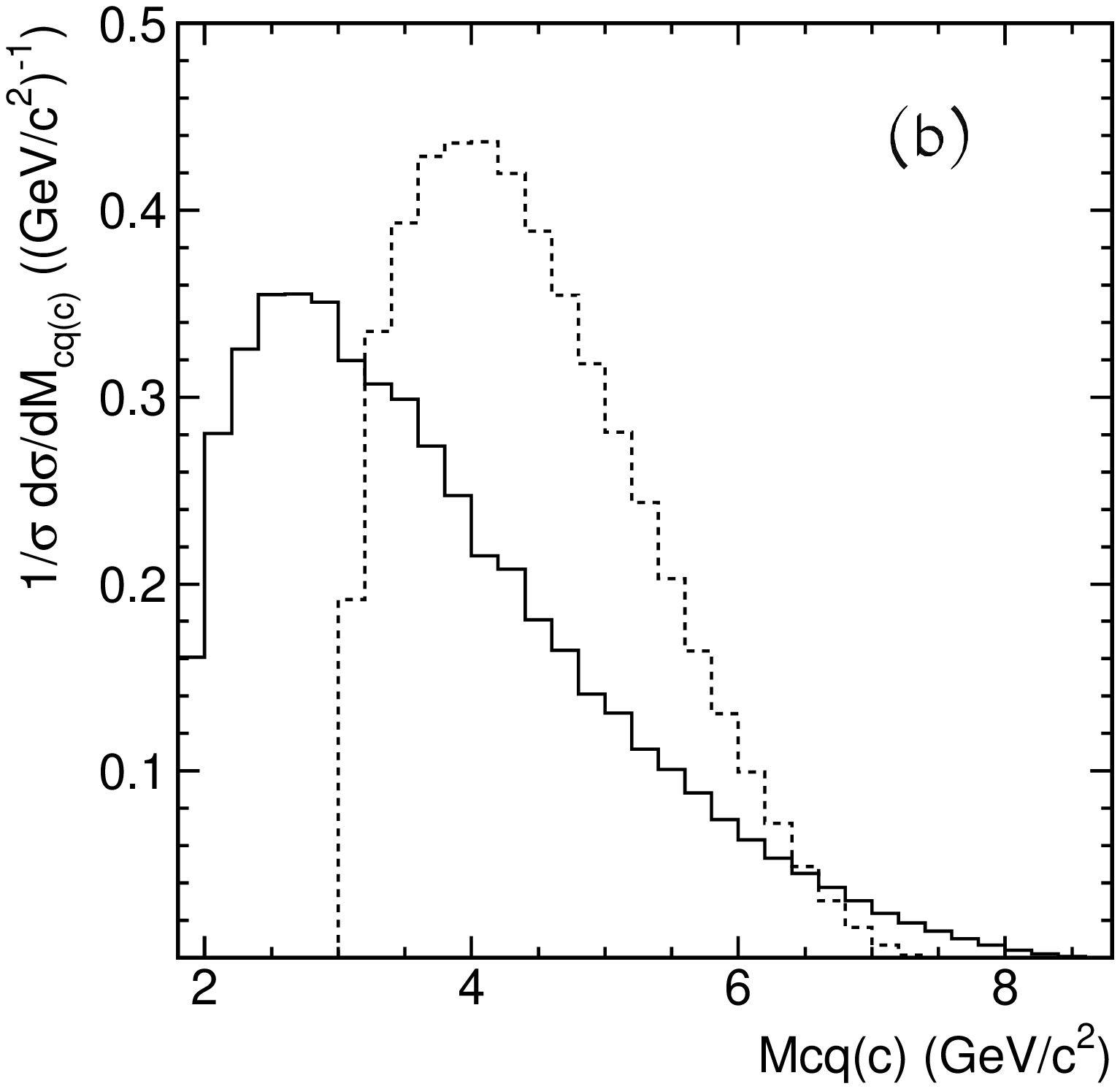, width=8cm}
\caption{(a) The dependence of cross section on quark mass for
$e^+e^- \to c \bar c q \bar q$ process (solid line) and 
$e^+e^- \to c \bar c c \bar c$ process (dashed line).
(b) The normalized distribution $1/\sigma~ d\sigma/dM$ for $e^+e^- \to 
c\bar c q\bar q$ process (solid line)
and $e^+e^- \to c\bar c c \bar c$ process (dashed line).}
\label{xsec}
\end{center}
\end{figure}

Before investigating the hadronization effects induced by
different colour connections, we discuss the theoretical uncertainties 
caused by quark mass, parton shower, etc. For charm quark mass, it is 
general to take the value between $1.2 GeV/c^2$ to $1.8 GeV/c^2$,
 and for light quark
mass,
the same value as that in Pythia is adopted. 
For $e^+e^- \to c\bar{c} q\bar{q}$ ($q=u,d$) process, we calculate
its cross section $\sigma(m_q)$ with  respect to light quark mass $m_q$.
At the same time, we investigate the production cross section
$\sigma(m_c)$ of $e^+e^- \to c \bar c c \bar c$.
The ratios for $\sigma(m_i)/\sigma(m_{i0})$($i=q,\, c$) 
with $m_{q0}=0.33 GeV/c^2$ and $m_{c0}=1.5 GeV/c^2$ are shown in
fig. \ref{xsec} (a). 
One can find that the cross section for $e^+e^- \to c \bar c q \bar q $ 
process is not quite sensitive to the light quark 
mass with charm mass fixed.
This implies that
the small light quark mass does not introduce  any
more dangerous singularities.
The differential distributions
$1/\sigma\, d\sigma/dM_{cq/cc}$ are  displayed in 
fig. \ref{xsec} (b).
It is easy to find
that the shape of the distributions 
for $e^+e^- \to c \bar c q \bar q$ and that for
$e^+ e^- \to c \bar c c \bar c$ (fig. \ref{xsec} (b))
are quite similar, and 
 a similar study on doubly charm diquark fragmentation for 
four charm process might also be  plausible. 
We also compare our results for the cross section of 
$e^+e^- \to c\bar{c} c\bar{c}$ with those given in \cite{jlee}, 
and find completely agreement.
The cross section for four quark production in $e^+e^-$ annihilation
at high energies (eg., LEP1 and LEP2) may be affected by the gluon 
radiation, which is simulated by parton shower process in Pythia.
Fortunately, at B factory energies, $\sqrt{s}\sim 10 GeV$, the
largest energy of each quark is about $5 GeV$. If it could 
radiate gluons further, its virtuality will quickly reach 
the lower limit that perturbative QCD can be applied safely. 
As a result, the parton shower process
at B factory energies would not play an important r\^{o}le as that at 
LEP energies. When one investigates physics for
the $e^+e^- \to h's$ process
at LEP or even higher energies, the parton shower process must be
taken into account. Unfortunately, till now, it is still unclear
how to describe the gluon radiation from diquark pair in $e^+e^-$
annihilation. As the first step of the investigation,  
we concentrate on the physics at the B factory energy, where the influence
of gluon  radiation seems to be small
and could be neglected. 

{\large \it Hadronization}~~
For the $e^+e^- \to c \bar c q \bar q$ process, 
when $m_1$ and  $m_2$ 
near the mass threshold of diquark,
 we employ Pythia to investigate the  
differences in  final hadron states corresponding  to  two
kinds of colour connections. 
One is that  $cq$ and $\bar{c}\bar{q}$ form diquark-antidiquark pair, 
and  hadronize altogether.
In the picture of Lund model,
diquark and antidiquark are in  $3^*$ and $3$ colour state,
respectively. As a result, one colour string will be stretched 
between them, and the string will fragment into hadrons at the end.
The other case is that $c\bar q $ and $\bar c q$ form two 
colour singlet clusters (strings). These two strings will 
fragment into hadrons independently.
This hadronization picture  is widely adopted in the popular
event generators, and hereafter we denote it  
as '2-string fragmentation'.
The hadronization of these two cases can be treated by the 
corresponding  subroutines in Pythia. 
For simplicity, we fix the 
invariant masses  of the 
(anti-)quark pair $cq$ ($\bar{c}\bar{q}$) to be the
(anti)diquark  mass used in Pythia.
The results obtained under this simplification  
can be considered as a good approximation of the average
over a reasonable (hence small) range of 
$cq$/$\overline{cq}$ cluster mass.
It is straightforward to adopt that the quantities we calculate in 
this section is not sensitive to the cluster mass in a small range around
the diquark mass set by Pythia.
By a similar argument, we also assume the inner orbital angular momentum 
of the $cq$ ($\overline{cq}$) cluster to be 0.
However, for treating the second kind of colour connection, 
the inner relative movements between $c$ ($\bar c$) and $q$ ($ \bar q$) 
are still not fixed and generated with 
the relative weight given by the differential cross section at
parton  level discussed above.

\begin{figure}
\psfig{figure=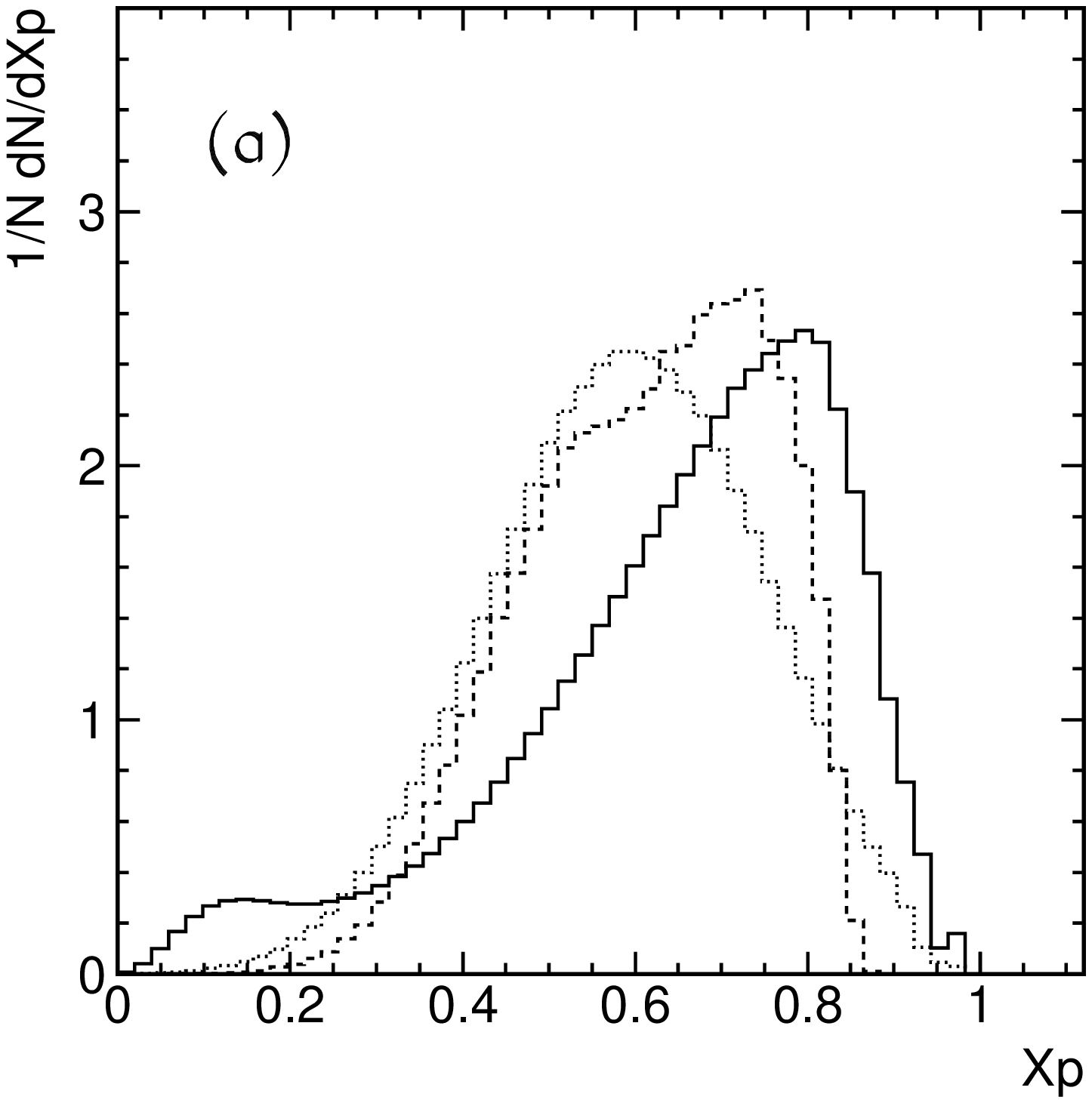, width=8cm} \psfig{figure=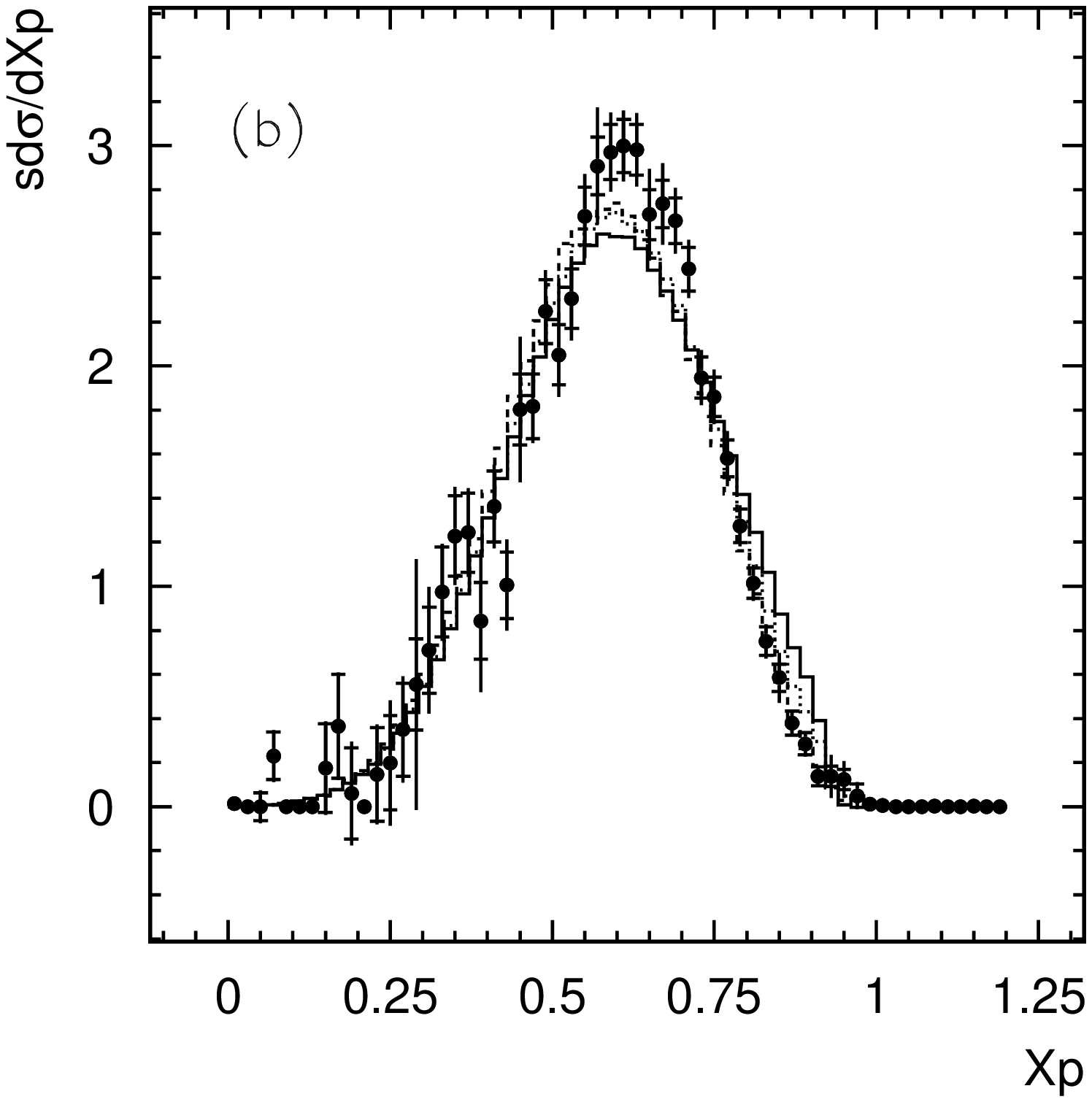, width=8cm}
\caption{(a) $X_p$ ($=\frac{p}{p_{max}}$, 
with $p_{max}=\sqrt{\frac{s}{4}-m^2}$) 
spectra of 
 $\Lambda_c$ at $\sqrt{s}=10.52 GeV$, from diquark fragmentation
(solid line), 2-string fragmentation (dashed line) and 
$e^+e^- \to c \bar c \to \Lambda_c +X$ (dotted line). 
(b) Comparing with data \cite{Seuster:2005tr}.
The histograms
correspond to three different baryon production
 scenarios in Pythia:
solid line for 'diquark', dotted line for 'simple popcorn' ,  dashed
line for 'advanced popcorn'.}
\label{zspectrum}
\end{figure}

In the diquark pair fragmentation process, it 
is straightforward that diquark tends to fragment  into   baryon, which is
in general the leading particle. If the charm baryons from the diquark 
fragmentation are detected, their momentum spectra should be   harder 
than those from  
the 2-string fragmentation,
or those from the general 
$e^+e^- \to c \bar c \to h's$ process. This feature can be displayed 
 by the momentum (fraction) spectra 
of  $\Lambda_c$ at   $\sqrt{s}=10.52 GeV$ (fig. \ref{zspectrum} (a)).
Recently, the momentum spectra
of various charm hadrons 
have been measured by BELLE Collaboration\cite{Seuster:2005tr}.
To make full use of the available data and to identify
whether the diquark fragmentation events exist or not,
we must make clear to what extent Pythia can reproduce the data. 
For charm meson production, we find 
that the predictions of Pythia agree to the available 
data quite well (also see \cite{cleo}).
For our aim, we concentrate on the investigation of 
charm baryon (eg., $\Lambda_c$) 
production. In Pythia, there are three 
scenarios to describe baryon production, 
i.e., diquark, simple popcorn  
and advanced popcorn. 
Their predictions can be 
consistent with each other provided  that the relevant parameters are tuned 
(fig. \ref{zspectrum} (b), the parameters  all take default values except 
$PARJ(1)=0.2$, $PARJ(18)=0.19$ and $PARF(192)=0$ for advanced popcorn). 
From fig. \ref{zspectrum} (b), one can find 
that the distributions of
$\Lambda_c$ predicted by standard Pythia 
are slightly softer than the data, while the momentum spectrum obtained 
by diquark fragmentation is slightly harder. 
It is also interesting to notice that
for the diquark fragmentation process, 
the small momentum region is slightly enhanced for the mass effect
(fig. \ref{zspectrum} (a)).
This phenomenon is also indicated  by the data
shown in fig. \ref{zspectrum}(b), though the 
statistical error is too large to give a definite conclusion.
Because of the phase space limit, the
probability of diquark pair  production may not be large.
As a result, the predictions for $\Lambda_c$ production including the
diquark pair production will not conflict with the available data,
and a portion of diquark fragmentation events is favoured. 

\begin{figure}
\psfig{figure=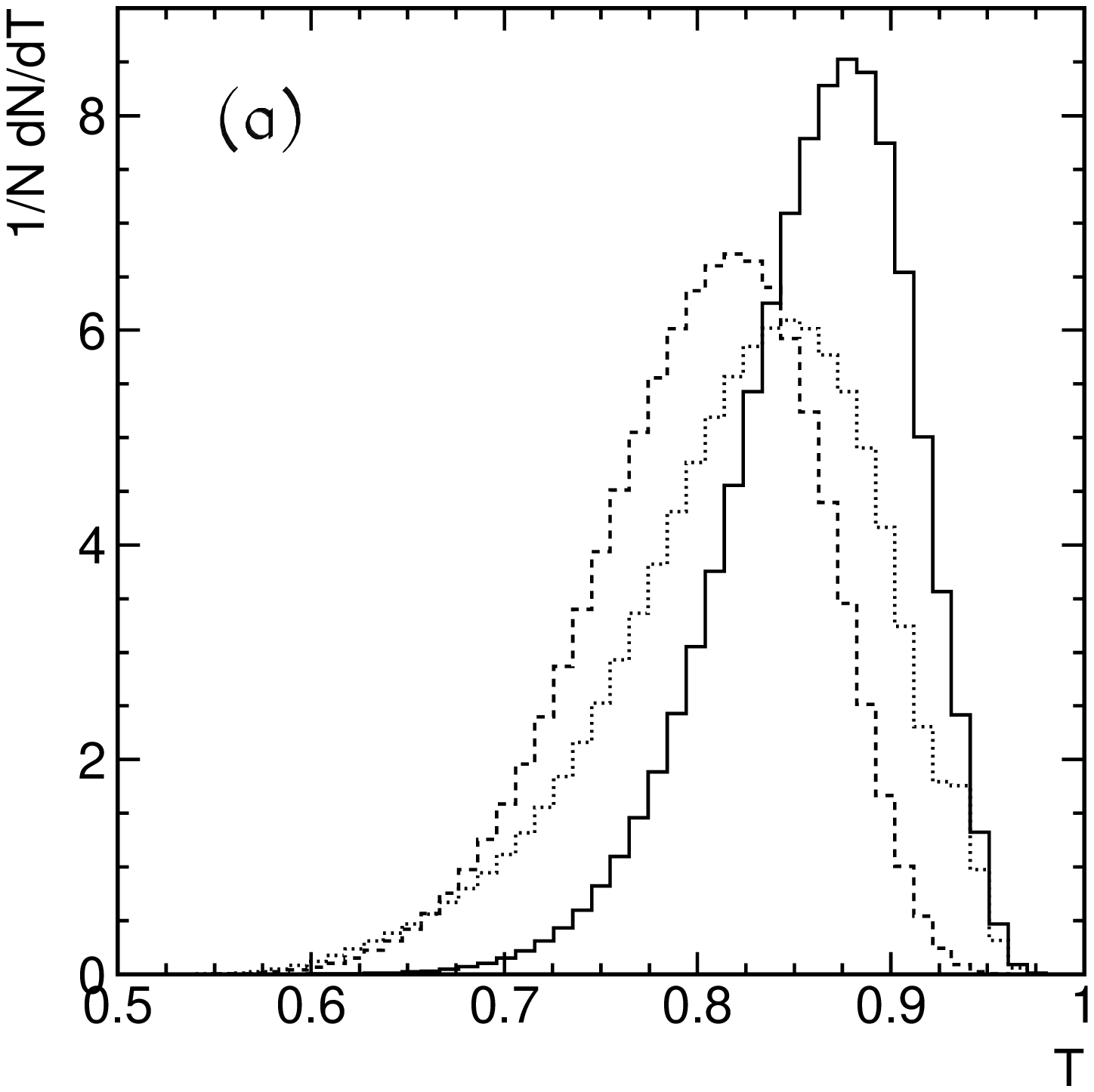,width=8cm} \psfig{figure=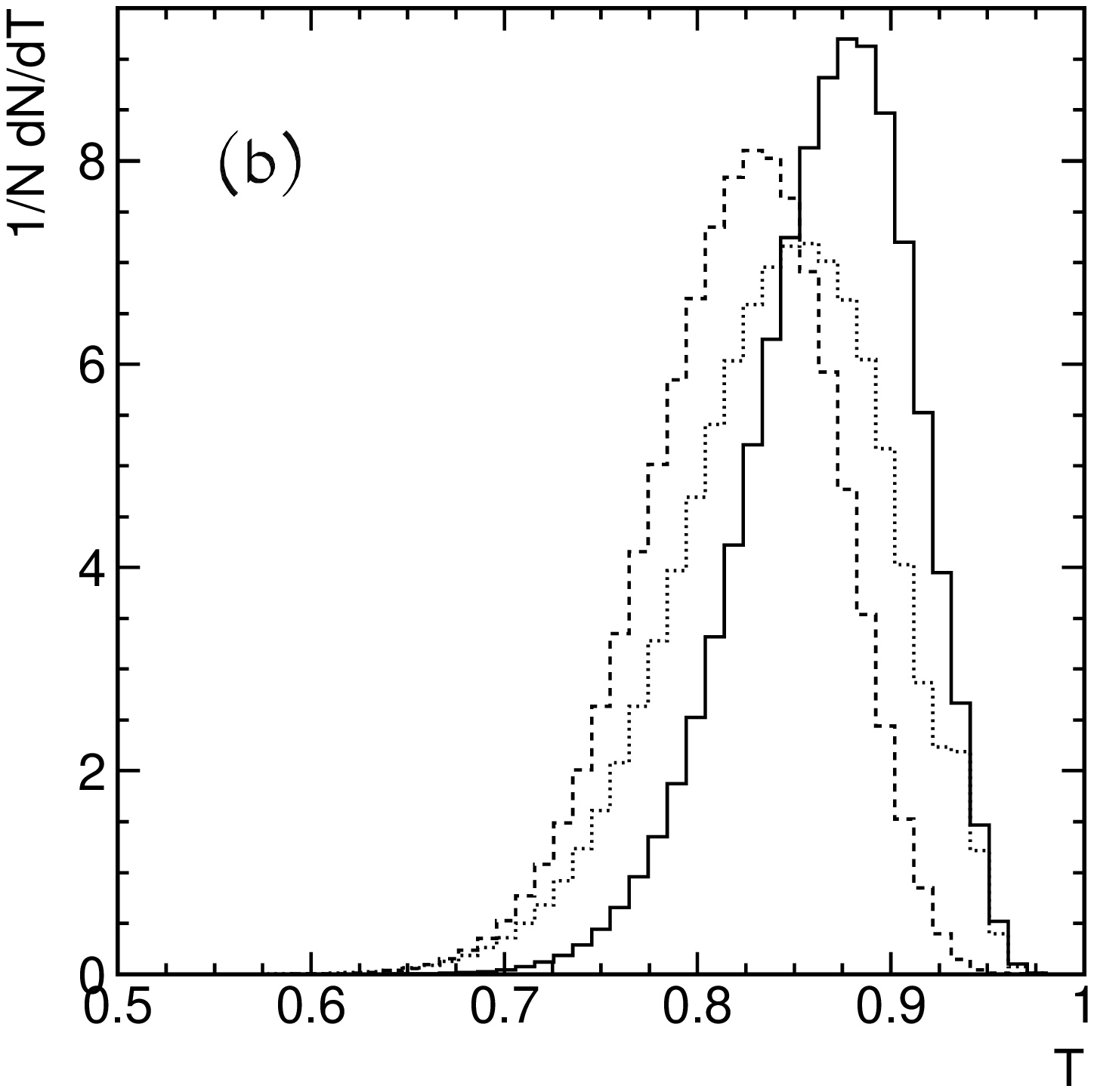,width=8cm}
\caption{Thrust distributions  of three cases at $\sqrt{s}=10.52 GeV$. 
(a)  Diquark fragmentation  (solid line), 
2-string fragmentation  (dashed line)
 and $e^+e^- \to c \bar c \to h's$ (dotted line). (b) Same as 
(a) except that  we select 2-jet events, with JADE Algorithm 
and $y_{cut}=0.12$. }
\label{2jet}
\end{figure}

For further investigation,  
we calculate thrust distribution $1/N~ dN/dT$
for the diquark fragmentation, 2-string fragmentation and 
$e^+e^- \to c\bar{c} \to h's$ process (fig. \ref{2jet}(a)).
One can find that the thrust distribution 
of the diquark fragmentation 
is much 'thinner' than those of the other cases (fig. \ref{2jet}(a)).
Aware  that the diquark fragmentation events are 
mostly 2-jet like, we  compare the thrust distributions 
of 2-jet events for the three cases.
We use the JADE algorithm with $y_{cut}=0.12$. In this case, around $90\%$
diquark fragmentation events are 2-jet. 
The properties  are similar to those without jet 
constrain (fig. \ref{2jet}(b)). 
According to  these discussions, further experimental 
investigation of  the colour connections shown in eq. (\ref{diqcl})
can be done by selecting 
2-jet charm  events at the B factory, and choosing those with 
large thrust (say, larger than 0.85). 

It is natural  that in the fragmentation of
diquark-antidiquark pair, a
baryon is produced together with an antibaryon.  
That means,  the correlation between charm and anti-charm 
baryons in the diquark pair fragmentation is  much 
stronger than that in the single charm quark 
fragmentation (either the 2-string fragmentation or the general 
$e^+e^- \to c \bar c \to h's $).
We know that, in the latter two cases, the production
probability for charm meson
is much larger than that for charm baryon.
This phenomenon can be illustrated by the baryon  ($B$) antibaryon ($\bar B$) 
correlation, defined as, 
\begin{equation}
A(B \bar B )=
\frac{2 N(B \bar B)}{N(B)+N(\bar B)},
\end{equation}
where  $N(B \bar B)$, $N_B$ and $N_{\bar{B}}$ respectively 
denote the probability for finding
$B\bar{B}$ pair, $B$ and $\bar{B}$.
We take the correlation of 
$\Lambda_c$ and $\bar \Lambda_c$ 
as an example in Table \ref{corr}. 
One can notice that the correlation in the diquark pair fragmentation process
is stronger than the other cases by about 10 times. 
This correlation is not sensitive to the gluon radiation of the diquark pair
 so we also give
the results at $Z^0$ pole.
This quantity
could be a good probe for the diquark pair fragmentation.
The results for other charm baryons,  such as  $\Sigma_c$ , $\Xi_c$ 
and the corresponding spin-3/2 particles are similar.

\begin{table}
\vspace{1cm}
\begin{center}
\begin{tabular}{|l||c|c|c|}  \hline\hline
{\em $\sqrt{s}$}&
{\em $e^+e^- \to c \bar c \to h's$}&
{\em Diquark Fragmentation }&
{\em 2-string} \\ \hline \hline
10.52GeV& 0.057    &0.76   &0.044\\ \hline
91.19GeV&0.081  &0.72  &0.081 \\\hline
\end{tabular}
\caption{$\Lambda_c\bar{\Lambda}_c$ flavor correlations.}
\label{corr}
\end{center}
\end{table}



To summarize, we investigate  the  hadronization effects induced by
different colour connections of four quark 
system (eg., $c\bar{c} q\bar{q}$) in $e^+e^-$ annihilation. 
It is interesting to point out that besides the normal 
colour structure with two colour 
singlets $c\bar{q}$ and $q\bar{c}$ which fragment
 into hadrons independently in the popular models,
there exists another kind of colour structure, 
as shown in eq. (\ref{diqcl}),
which has to be taken into account in some cases.
For that colour connection, needless to 
say, the hadronization of these four quarks should be treated  as a
whole system,
i.e., they interact among each other during the hadronization process. 
In this letter, considering an extreme limit case 
that $cq$ and $\bar{c}\bar{q}$
form diquark-antidiquark pair, we propose a toy model based on   Pythia 
to describe the hadronization of them. It is found that  
the results originated from diquark fragmentation 
are significantly different from those of the ordinary colour connections.
However because of the phase space limit, 
the hadronization effect from diquark pair fragmentation 
is not very large. To understand the hadronization mechanism 
induced by different
colour connections of final  partons produced perturbatively, 
further theoretical and experimental investigations
are still necessary.


\section*{Acknowledgments}
This work is supported in part by NSFC, NCET of MoE, P. R. China, 
and Huo YingDong Foundation.
The authors thank all of the members  
in  Theoretical Particle Physics Group of Shandong 
University for their helpful discussions. The Author (Si) would like
to thank the hospitality of the Phenomenology research group
of Department of Physics, University of Wisconsin-Madison during his visit.

\end{document}